\documentstyle[12pt]{article}
\newcommand{\be}{\begin{equation}}
\newcommand{\ee}{\end{equation}}
\newcommand{\bdis}{\begin{displaymath}}
\newcommand{\edis}{\end{displaymath}}

\newcommand{\bom}{\bf{\omega}}
\newcommand{\bnabla}{\bf{\nabla}}

\newcommand{\bv}{\bf{v}}
\newcommand{\bk}{\bf{k}}

\title{Time-Reversible Dynamical Systems for Turbulence}
\author{ L. Biferale$^1$, D. Pierotti$^2$ and A. Vulpiani$^3$}
\begin{document}
\maketitle
\centerline{$^{1}$  Dipartimento di Fisica, Universit\`{a} di Tor Vergata}
\centerline{Via della Ricerca Scientifica 1, I-00133 Roma, Italy and}
\centerline{Istituto Nazionale di Fisica della Materia, unit\`a di Tor Vergata}
\centerline{$^{2}$ Dipartimento di Fisica, Universit\`{a}
dell' Aquila }
\centerline{Via Vetoio 1, I-67010 Coppito, L'Aquila, Italy and}
\centerline{Istituto Nazionale di Fisica della Materia, unit\`a dell'Aquila}
\centerline{$^{3}$  Dipartimento di Fisica, Universit\`{a} di  Roma
"la Sapienza"}
\centerline{Piazzale Aldo Moro 5, I-00185 Roma, Italy and}
\centerline{Istituto Nazionale di Fisica della Materia, unit\`a de "La 
Sapienza"}
\date{ }
\medskip

\begin{abstract}
{\it Dynamical Ensemble Equivalence} between hydrodynamic
dissipative equations and suitable time-reversible dynamical systems
has been investigated
in a class of dynamical systems for turbulence.\\
The reversible dynamics is obtained
from the original dissipative equations by imposing a 
global constraint.\\
We find that, by increasing the input energy,
the system changes
from an equilibrium state to
a non-equilibrium stationary state in which an energy cascade, with
the same statistical properties 
of the original system, is clearly detected.
\end{abstract}
\medskip
PACS number 47.27.Jv, 47.90.+a, 05.45.+b
\medskip

\centerline{\bf To the memory of Giovanni Paladin.}  
\newpage
 
\section{Introduction}

One of the most important open problem in  classical physics
is the understanding the statistical features of  fully developed
turbulence (FDT).

A fully developed turbulent flow is a dissipative
system described by the Navier-Stokes 
(NS) equations in the
limit of high Reynolds numbers ($Re$).  

On one hand, direct numerical simulations of turbulent flows  are strongly 
limited
 due to the huge amount of excited degrees of freedom:
a simple argument due to   Landau, 
 shows that the number of degrees of freedom, 
 which should be taken into account for a correct description
  of a turbulent flow, increases as $Re^{9/4}$.

On the other hand, analytical attempts to derive
the multi-points velocity probability distribution function have repeatedly
failed due to the strong  coupling regime and due to the highly
non-gaussian probability distribution functions
(PDF) developed at  small scales  by the velocity field \cite{frisch}. 
Phenomenological approaches, or simplified dynamical and deterministic systems, 
have been
therefore often  used for studying the  mechanisms generating  the 
turbulent energy cascade. 

From the analytical point of view, the main obstacle to  the
possibility of performing the $0$-th step toward a theory of
turbulence is certainly connected to the strong dissipative  and
far-from-equilibrium character of 3d
turbulent flows. In 2d turbulence, where 
energy is almost not dissipated at all, some analytical tools
based on quasi-equilibrium statistical ansatz have indeed been developed
\cite{sommeria}.

Strongly chaotic dynamical systems as the Anosov systems are 
the only cases  where, although being  still dissipative and chaotic, analytical
tools have been developed with relative success, at least in the case of
low-dimensions \cite{beck}.

Recently \cite{galla_cohen}, Gallavotti and Cohen
proposed that a chaotic high-dimensional dynamical
system in a stationary state can be regarded as a smooth dynamical system with 
a transitive Axiom-A global attractor or, if it is time-reversible,
as a smooth transitive Anosov system, as far as macroscopic
properties are concerned. This is the so-called {\it Chaotic Hypothesis}.

This hypothesis, as the ergodic hypothesis, can be proved only in very
particular systems. Nevertheless, it is interesting to analyze
some of its consequences.

Likewise, Gallavotti \cite{galla1,galla2} 
conjectured a {\it Dynamical Ensemble 
Equivalence} between some dissipative systems (in this paper we will
refer only to 3d Navier-Stokes equations) 
and their {\it non-equilibrium but time-reversible} formulation.
Here, equivalent must be meant 
that the averages of local variables in the two
systems, i.e. the original one and
its time-reversible formulation, 
are the same, in a suitable limit. For hydrodynamic
dissipative systems this limit is that of FDT, 
i.e. $Re\rightarrow\infty$.

The reversible dynamics is obtained from the original dissipative 
equations by
imposing a constraint, such as to keep constant in time
those macroscopic observables (as the total energy) 
which would have only had 
stationary
averages in the original systems.

Having a time-reversible system and applying the chaotic hypothesis,
some large-deviation properties of the fluctuation
of the entropy-production rate in the system can be proved \cite{galla2}. 

In Navier-Stokes equations, the reversible dynamics is achieved
by introducing a sort of "eddy viscosity" which removes the 
input energy with perfect efficiency.  Viscosity  becomes non-positive
defined and strongly correlated with the large-scale flow where energy is 
injected.

As far as we know, the idea of reversible NS equations
was introduced for the first time by She and Jackson \cite{she_jac}.
They did not exploit global constraints, but
imposed that the energy contained in each ``momentum shell'' was
constant.

Let us also mention that a constraint of constant energy has been implemented
by using 
the Kraichnan's eddy-viscosity 
parameterization \cite{kraic_eddy} in low-resolution large-eddies simulations
of NS equations in \cite{brisco}. 
With such a parameterization of the viscosity one has very weak fluctuations
of the energy (less than 1
of high-resolution numerical simulations.

From our point of view, the interest in models with global
constraints stems from
the possibility to describe a global macroscopic dissipative and irreversible 
physics starting from a deterministic reversible dynamics. 
The approach can be seen as a bridge from microscopic reversible 
dynamics to macroscopic irreversible dynamics and, more interesting,
a possible systematic tool  
for going  with continuity from a pure-equilibrium and conservative
systems to a strong dissipative and far-from-equilibrium time evolution.

In this paper, we investigate these ideas in Shell Models, i.e.  
 a class of simplified dynamical 
systems for turbulence (for a recent review see \cite{bjpp}, for a 
tutorial introduction see \cite{kadanoff}). 
In particular, we will analyse in details the smooth transition from
the equilibrium system at zero viscosity and zero external forcing
to a (formally)-reversible systems which possess anyway a 
non-equilibrium flux of energy from large to small scales.

The paper is organised as follows. In section 2 we briefly 
review the statistical mechanics of a perfect fluid and
the ideas presented by Gallavotti in \cite{galla_cohen,galla1,galla2}
concentrating only to the case of Hydrodynamical systems (Navier-Stokes eqs.).
In section 3 we discuss Shell Models philosophy  and we describe
 Gallavotti's implementation to our case. 
In section 4 we present our numerical results. Conclusions follow in section 5.

\section{Equilibrium and Non-equilibrium Statistical Mechanics}

In a $3 D$ perfect fluid, i.e. with vanishing external forcing $\nu=0$,
the evolution of the velocity field 
is given by the Euler equations which 
 conserve  two quadratic functional, the 
kinetic energy and the helicity:
\begin{equation}
E = \frac{1}{2} \, < v^2 > \qquad H= {1\over 2} \, <\bv \cdot \bom >,
\label{1}
\end{equation}
where $\bom =\bnabla \, \times \, \bv$ is the vorticity. 
  In this case,  it is possible to construct a statistical mechanics
  as for a gas:
 by   using the conservation laws 
 and the conservation of the volume in phase space
 one obtains a gaussian distribution.

For simplicity let us start by neglecting the helicity conservation. 
To be explicit, let 
  us consider an  incompressible inviscid fluid in a 
 cube with periodic boundary conditions, so that the velocity field
 can be expanded in Fourier series as
\begin{equation}
v_j({\bf x})=L^{-3/2} \sum_k e^{i {\bf k} \cdot {\bf x}} v_j({\bf k}) 
\label{2}
\end{equation}
with ${\bf k}=2 \pi {\bf n}/L$ and ${\bf n}=(n_1,n_2,n_3)$, where $n_i$
 are integers. The variables $v_j(\bk)$ are not completely 
 independent, since from the 
 incompressibility condition and the fact 
that ${\bf v}({\bf x})$ is real, it follows that
$$
{\bf k} \cdot {\bf v}({\bf k})=0 \qquad {\rm and} \qquad 
{\bf v}({\bf k})={\bf v}^*(-{\bf k})
\label{3}
$$
In any case, it is straightforward to introduce a new set of independent 
variables $X_a$, where now $a$ labels the spatial component and the 
wave vector. By using an ultraviolet truncation, 
${\bf v}({\bf k})=0$ for $k > k_{\rm max}$, and 
by introducing (\ref{3}) in the Euler equations one obtains a set of 
ordinary differential equations (ODEs)  with the structure
$$
{dX_a \over dt}= \sum_{b,c} M_{abc} \, X_b\, X_c
\label{4}
$$
where $M_{abc}=M_{acb}$ and $M_{abc}+M_{bca}+M_{cab}=0$
 with $a=1,\cdots,N \sim k_{max}^3$.
We stress the fact that the ultraviolet truncation is necessary
in order to avoid the infinite energy problems of  classical field theory.

 It is easy to verify that \ref{4}  preserves the volume
 in the phase space as well as the energy, namely
$$
\sum_a {\partial \over \partial X_a} \, \left( {dX_a \over dt} \right)
 =0 \qquad {\rm and} \qquad {dE \over dt}= {1\over 2}{d\over dt}
\sum X_a^2=0
\label{5}
$$
These conservation laws are sufficient to construct the probability 
distribution 
 of the variables $\{X_a\}$  \cite{kraiMon80}: 
 using the ergodic hypothesis,  one 
 obtains the microcanonical probability measure
$$
P_m(\{X_a\})\sim\delta\left( {1\over2}\sum_aX_a^2-E \right)
\label{6}
$$
It is well known that,  in the limit $N \to \infty$, 
this  is equivalent to the canonical measure
$$
P_c(\{X_a\}) \sim \exp - \left({
 \beta \over 2} \sum_a X_a^2 \right) 
\label{7}
$$
where the Lagrange multiplier $\beta$ satisfies the relation
$$
<X_a^2>={2\, E \over N}=\beta^{-1}
\label{8}
$$
In two dimensions, the helicity $H \equiv 0$ and 
  there exists a second conserved quantity, the 
enstrophy
$$
\Omega={1\over 2} \int \omega^2 \, d^2x
\label{9}
$$
which is the mean square vorticity. In terms of the $X$ variables,
 it can be written as
$$
\Omega={1\over 2} \sum k_a^2 X_a^2
\label{10}
$$
As a consequence, the microcanonical probability measure in $2d$
 is
$$
P_m(\{X_a\})\sim
 \delta\left( {1\over2}\sum_a X_a^2-E \right) \ 
\delta\left( {1\over2}\sum_a k_a^2 X_a^2-\Omega \right)
\label{11}
$$
and the corresponding canonical measure is 
$$
P_c(\{X_a\})\sim \exp - \left(
 {\beta_1 \over 2} \sum_a X_a^2 + 
 {\beta_2 \over 2} \sum_a k_a^2 X_a^2 \right) 
\label{12}
$$
where the Lagrange multipliers satisfy the relation
$$
<X_a^2>={1\over \beta_{1} +\beta_2 \, k_a^2 }
\label{13}
$$
In 3-d one can repeat a similar argument, taking into account the helicity
conservation. In this case, being the helicity non positive defined,
one as to assume suitable constraints for the generalized temperature
related to the helicity \cite{kraich}.

The above results are, both in $2d$ and in $3d$, 
 well reproduced by numerical simulations \cite{kraiMon80}.

The limit $\nu \to 0$
 (equivalent to $Re \to \infty$) is singular and cannot be
 interchanged with the limit $N \to \infty$. Therefore, the statistical 
mechanics 
 of an inviscid fluid has a quite limited relevance on the 
behaviour of the Navier-Stokes equations at high Reynolds number.
Recently some authors proposed the use of conservative statistical
mechanics to justify some behaviours of real fluids, e.g the Jupiter's red spot
and the emergence of organized structures 
\cite{RobertSommeria991,Miller1992,Pasmanter1994}. 
The applicability of this approach is limited to some particular 
quasi equilibrium two-dimensional situations.

On the other hand, both from phenomenological arguments and experimental
results, we know that the statistical mechanics of fully developed turbulence
has peculiarities rather different from the usual statistical mechanics
of conservative systems.
In the limit of FDT the energy fluctuates around its mean value and
in addition one has an energy cascade from  large to small
scales.

The turbulence is described by a dissipative
system (essentially a high dimensional truncation of the Navier-Stokes 
equations with
 $|{\bf k}| < k_{max}=O(Re^{3/4}$) in which the volume in the phase
space is not conserved. Let us stress again that  the two limits $\nu \to 0$ 
and $k_{max} \to \infty$ must be take in a suitable way in order to obtain
the correct physical result for the turbulence.
If one wants that the mean energy dissipation is $O(1)$ in the limit
$Re \to \infty$ one has to take $k_{max}>O(Re^{3/4})$.

In order to have a statistical  stationary state one needs two basic
ingredients: a `friction' mechanism and a coupling with an
external forcing or `reservoire'. A typical example of statistical 
 stationary state is given by conductive systems where an external electric
field and a friction mechanism, mimicking the electrical resistivity, 
leads to a macroscopic steady current.

Recently time-reversible and conservative systems have been introduced
in the issue of the non-equilibrium statistical mechanics of stationary
 state.
For sake of self consistency  we recall one of the simplest system of this
class.
Let us consider $N$ independent particles of mass $m$,
with coordinates and momenta ${\bf q}_i$ and  ${\bf p}_i$
respectively, on a square domain (in $2$ or $3$ dimensions)
 with periodic boundary condition for the variables 
${\bf q}_{i}$. Using a suitable  external potential $V({\bf q})$ we can
 mimic the elastic scattering with rigid obstacles in order to have
basically  a "Lorentz gas".
The equation of motion are:
\begin{eqnarray}
&&\frac{d {\bf q}_{i}}{dt}=\frac{1}{m}{\bf p}_{i}
\label{eq:loren1}\\
&&\frac{d {\bf p}_{i}}{dt}=-\frac{\partial V}{\partial {\bf q}_{i}}
\label{eq:loren2}
\end{eqnarray}
The system is chaotic and one can expect the usual microcanonical
distribution. Of course there is no net current.
In order to have a current in the $x$ direction it is necessary to
add in the eq. (\ref{eq:loren2}) 
a term $E {\bf e}_{1}$, where ${\bf e}_{1}$ is the versor in the $x$ direction.
At the same time, if one wants
to focus on stationary aspects, some energy-loosing mechanism must
be added.  Standard phenomenology would suggest the insertion of
a viscous irreversible term of the form $ -\alpha {\bf p}_{i}$ in the equation
of motion governing the evolution of momenta. 
In this way,  one is naturally lead to a stationary state.

Recently, in \cite{bonetto}, the idea of mimicking this behaviour
by means of an exactly conservative and reversible physics has been
proposed by using instead of a constant viscous coefficient $\alpha$,
a perfect energy-sink, correlated with all scales and able to reabsorb
instantaneously all excess of energy injected by the forcing term
in the system. This ideal viscosity must acquire an explicit time-dependency
and works out as a Lagrange multiplier  such as the total energy
is an invariant of motion. Being the forcing mechanism not-positive defined,
also the ideal viscosity will be not-positive defined. 

The system now develops a net current and all the phenomenology
of a dissipative physics. The natural question which arises is
how much the original dynamics is preserved by this 
very-strong perturbation of the equation of motion. 

In \cite{bonetto} some numerical simulation of eqs. 
(\ref{eq:loren1})-(\ref{eq:loren2}) have been performed
showing that some of  the main signature of the original physics are still
present in the modified model with the advantage,  in the latter, that 
also some analytical investigation can be carried out. 
In particular the most important consequence of the chaotic hypothesis
is the {\it fluctuation theorem} that is an exact parameterless prediction.
This theorem concern the probability distribution function of the 
contraction rate of of the attractor surface element
(for a detailed discussion see \cite{galla2}). Let us note
that is very difficult, save for very particular systems,
to test this prediction as the attracting sets usually are unknown.\\
In the next section we will investigate a similar problematic in 
a class of dynamical systems for turbulent flows, called Shell-Models.
In particular, we want to understand how much freedom is allowed in
the choice of a viscous-modelization without perturbing too much
the main physical framework, and/or quantifying the aspects of the
perturbation, eventually.\\ 
The goal consists in having a reversible dynamics showing in some limit
(to be defined) the same physics of a turbulent dissipative flow. 

\section{Time-reversible Shell Models}

One of the most intriguing problems in 3 dimensional turbulence 
is related to the understanding of the non linear 
dynamical mechanism triggering 
and supporting the energy 
cascade from large to small scales. Following the
Richardson scenario that energy should be transferred downwards in scales, 
Kolmogorov \cite{K} (K41) postulated that the energy cascade should
follow a  self-similar and homogeneous process entirely dependent
on the energy transfer rate, $\epsilon$. This idea, with the assumption of 
local isotropy and universality of the small scales, 
eventually led  to a precise prediction:
\be
S_p(l) \equiv \langle(\delta v(l))^p\rangle=C_p 
\langle (\epsilon(l))^{p/3} \rangle l^{p/3} 
\label{eq.1}
\ee
where $C_p$ are constants and $\epsilon(l)$ is the coarse-grained
energy dissipation over a scale $l$ supposed to be in the .
 inertial range, i.e. much smaller than the integral scale 
and much larger than the viscous dissipation cutoff.
If $S_p(l)\sim l^{\zeta(p)}$ and
 $\langle \epsilon^p(l)\rangle\sim l^{\tau(p)}$
then
\be
\zeta(p)=p/3+\tau(p/3)
\label{eq.2}
\ee

In K41 the $\epsilon(l)$ statistic   is assumed to  be 
$l$-independent, or $\tau(p)=0$, implying
$\zeta(p)=p/3, \forall p$, in particular $\zeta(2)=2/3$ or, equivalently,
the energy spectrum going as $k^{-5/3}$.
On the other hand, there are many experimental and numerical
 \cite{MS,BCTBS,BCBC,Kerr90} 
results telling  us that K41 scenario for homogeneous
and isotropic turbulence is quantitatively wronged. Strong intermittent
bursts in the energy transfer have been observed and non trivial $\tau(p)$
set of exponents measured. 

Shell models have demonstrated to be very useful for the understanding
of many properties connected to the non-linear turbulent energy
transfer \cite{G}-\cite{BK}.
The most popular shell model, the Gledzer-Ohkitani-Yamada 
(GOY) model (\cite{G}-\cite{BK}), has been shown to predict
scaling properties for $\zeta(p)$ (for a suitable  choice of the 
parameters) similar to what is found experimentally.  

The GOY model can be seen as a severe truncation of the 
Navier-Stokes equations: 
it retains only one complex mode $u_n$ as a representative 
of all Fourier modes in 
the shell of wave numbers $k$ between $k_n=k_02^n$ 
and $k_{n+1}$.

Dynamical equations 
have  the same qualitative structure of Navier-Stokes eqs., namely:
\be
  \frac{d}{dt} u_n = N_{n}[{\bf u}] -\nu k_{n}^{2} u_n +f_{n}
\ee
where $N[{\bf u}]$
 are the inertial nonlinear terms (see below), while $\nu$ is
the molecular viscosity and $f_{n}$ a suitable forcing term acting
only at large scale introduced in
order to reach a (statistical) stationary state.

The  choice of the nonlinear term is 
dictated from the  "locality assumption", i.e. 
only couplings with the nearest and next nearest shells are kept. In details
the final eqs. are:
\be
  \frac{d}{dt} u_n =i\, k_n \left(u^{*}_{n+1}u^{*}_{n+2} + 
  b u^{*}_{n+1}u^{*}_{n-1}
  +c u^{*}_{n-1}u^{*}_{n-2} \right) 
   -\nu k_n^2 u_n +\delta_{n,n_0}f,
\label{goy}
  \ee
  where the  the external forcing acts
on a large scale $n_0$ and $b$, $c$ are two free parameters,
but with the constraint $1+2b+4c=0$, used
for changing the "dimensionality" of the system  \cite{LKWB,BK}, 
a popular choice
which leads to  results  close to the 3d turbulent 
phenomenology is $b=-1/4, c=-1/8$. Let us stress that this choice of 
the parameters corresponds to have both energy and helicity conservation
for a shell scale ratio equal to 2 and that whenever one has these two
invariants, for any shell ratio, one has anomalous scaling
exponents.\\
At fixed molecular viscosity, $\nu$, the model develops 
a chaotic energy transfer
to the small scales, with intermittent burst and deviation from K41 in good 
qualitative and quantitative agreement 
with what observed in true turbulent flow.

The natural question which we would 
like to analyze in this paper is whether 
a reversible system obtained from the original dissipative equation by 
imposing a global constraint will allow us to reproduce the 
standard results and whether one can learn something of more about
the strong-dissipative and far-from equilibrium structure of the stationary
statistics. 

Therefore, following Gallavotti's suggestion 
we introduce a Lagrange multiplier
$\alpha[\bf{u}]$ 
such as the eqs. of motion (\ref{goy}) preserve the total energy for any time, 
namely:
\be
 \frac{d}{dt} u_n = N_n[{\bf u}] -\alpha[{\bf u}] k_n^2 u_n +f_{n}.
\label{goyvinc}
\ee
In order to have the total energy 
$E=1/2\sum|u_{n}|^{2}$  constant, one has to impose:
\be
\alpha[{\bf u}] = \frac{{\Re}(u_0 f)}{\sum_n k_n^2 |u_n|^2}
\ee

Let us comment that in \cite{galla1,galla2} different versions of 
reversible - hydrodynamical equations 
for a flow  have been proposed depending on which macroscopic
observables one fixes by mean of the lagrangian multiplier. For example,
equations similar to (\ref{goyvinc}) but with $\alpha[\bf{u}]$ chosen
such that the total energy dissipation is conserved  could in principle 
be used as well. 
In our view, guided from the phenomenological  behaviour of turbulent flows,
we believe that the only realistic constraint one can safely impose
to the equation is on the total energy. 
Constraining the total energy dissipation
would put too much weight on the small scales statistics and would kill
one of the most remarkable signature of turbulent flows: multifractal 
nature of energy dissipation. 

The goal of our study is to understand how the system move away from the
stationary and equilibrium state that one obtain when $f_{n}=0$
as soon as  some energy  pump and (perfect) energy sink are switched on 
$( f_{n} > 0)$.

\section{Numerical Results}

We first performed a benchmark numerical integration
of a standard irreversible and dissipative 
GOY model with fixed viscosity and forcing. This integration allows 
us to fix "physical realistic" values for the observables  of the 
reversible dynamics.  Numerical evolution was given  by a forth-order
Runge-Kutta algorithm,
for a  GOY model with $N=23$ shells and a constant
forcing on the first shell, $f=5\times10^{-3}\;(1+i)$.  
The integration time was several hundreds
 characteristic eddy turn-over times.
We measured the
structure functions and the average energy of the system.
Afterwards, we integrated the reversible dynamical system 
keeping
the total energy fixed to the mean value of the benchmark run.
We kept all the other parameters of the model equal to those
of the benchmark run, 
except for the value of the forcing which we
let vary in order to switch continuously from a 
conservative equilibrium dynamics ($f=0$) to 
a conservative non-equilibrium dynamics ($f >0$). \\
In fig. 1 we show the behaviour of 
functions:
\begin{equation}
\Sigma_{n,p}=\left\langle\left|{\Im}\left[
u_{n}u_{n+1}u_{n+2}+\frac{1}{4}
u_{n-1}u_{n}u_{n+2}\right]\right|^{p/3}\right\rangle,  
\end{equation}
for $p=2$ and
for different values of the external forcing.
$\Sigma_{n,p}$ represent the power $p/3$ of
energy flux from from shell $n$th to shell
$(n+1)$th divided by $k_{n}$. 
It is easy to show that  in the inertial range the energy flux must
be constant \cite{PBCFV}. Therefore, it is natural to quantify the 
statistical properties of the energy transfer by measuring
the scaling properties: $\Sigma_{n,p} \simeq k_{n}^{-\zeta_{p}}$.
One can observe that in the limit 
with  vanishing  forcing there is equipartition between  degrees
of freedom. 
In this case the viscous term 
is very low - the viscosity
is proportional to the value of the forcing -
and we have essentially a truncated-Euler  system with gaussian 
probability distributions
of the shell variables $u_{n}$.
When the forcing is
increased there appears two different scaling ranges.
In the first range
(small $k$'s)
it is clearly distinguishable an energy cascade. In the second range
(large $k$'s)
the energy is in equipartition among the degrees of freedom. Likewise
the probability distribution functions of  shells in equipartition
 have all the same functional non-gaussian form.
\\ 
The range in which
the energy cascade is observed is longer for higher forcing 
up to a critical forcing where the cascade range coincide with
the inertial range of the original GOY model.\\
For even higher forcing
the system falls in a 
stable fixed point in which all the energy is concentrated in the
first shell and all the other shell variables are zero.\\

We have checked that the  cascade  
range is not due to finite size effects
by performing a simulation 
with   a larger number of shells ($N=28$) and
keeping all the others parameters constant.
 
The same behaviour has 
been obtained in a model in which the dissipation term has 
been put only in the dissipative range (the last $7$ shells).

Likewise, we measured the
scaling exponents of $\Sigma_{n,p}$ in the energy-cascade range, using
the extended self similarity (ESS) in order to have
more accurate fits. In the ESS one measures the behaviour of the structure
functions of order $p$ versus the structure function of order $3$.
 In this way the scaling 
range is longer and measures of scaling exponents 
are more precise \cite{BCTBS}.
\\ 
In fig. 2 we plotted
the function of order $6$ and in fig. 3 we show
the behaviour of the scaling exponents compared to that of the GOY model.
There is a clear intermittent energy cascade.\\

On heuristic grounds, Gallavotti \cite{galla2}
made the conjecture that a dissipative system and its time-reversible
analog  should be equivalent (in a statistical sense) if 
the time scale by which the viscosity 
reaches its mean value is shorter
than the hydrodynamic time scales (i.e. the eddy turnover times).
In this case the  viscosity time-evolution 
would be  confused with its average.\\
Although we did not find a quantitative agreement
with the conjecture of Gallavotti the qualitative behaviour
is in the right direction.
We defined the characteristic time $\tau$ of $\alpha[\bf{u}]$ as 
the decaying time of the autocorrelation of its logarithm, that is
defined by:
\begin{equation}
C( \tau)=
\frac{<x(t+\tau)x(t)>-<x(t)>^{2}}
{<x(t)^{2}>-<x(t)>^{2}}
\end{equation}
where $x=\log(\alpha)$.
We considered the logarithm of the viscosity function 
because the function itself has very large fluctuations  
(several order of magnitude larger than its average)
and consequently the time average of its square has very long
convergence times. 
We have found that in the
case with the longer cascade range, i. e. in the system with
statistical properties closer to the original
GOY model, this characteristic time is shorter  
(see fig. 4) and consequently the time by which the viscosity reaches
its average is shorter. 
Moreover let us stress that in this case
$\alpha[\bf{u}]$ 
has a smaller mean square value, i.e. smaller fluctuations.

\section{Conclusions}
  The statistical mechanics of fully developed turbulence has
  features which are very different from those of the usual
  equilibrium statistical mechanics of Hamiltonian systems.
  Recently, Gallavotti proposed a dynamical ensemble equivalence
  between hydrodynamical dissipative system, e.g. the  Navier-Stokes
  equations, and time-reversible systems.

  In this paper, we introduce and study a  time-reversible dynamical 
  system obtained, from  a shell model for turbulence, changing the viscous
  term  such as the energy is conserved.
  At small forcing values the system has  statistical behaviours
  very close to those of a gas, i.e. energy equipartition and gaussian
  statistics. At increasing the forcing one has a non-equilibrium
  statistical stationary state with an energy cascade and anomalous
  scaling laws  similar to those observed in turbulence.

  The dynamical ensemble equivalence seems to be satisfied at least
  for the typical observables measurable in turbulent flows.

  A  relevant open problem remains for defining precisely
the class of constraints allowed for the {\it Dynamical 
Ensemble Equivalence} hypothesis to hold.

\section{Acknowledgement} 
This work was supported by INFN (Iniziativa Specifica {\it
Meccanica Statistica} FI11) and INFM (Progetto di Ricerca Avanzato 
{\it TURBO}).\\
We thank F. Bonetto and G. Gallavotti for many stimulating  discussions
and continuous encouragement. We also thank L.P. Kadanoff and
R. Pasmanter for  useful
discussions.

\newpage

\centerline{FIGURE CAPTIONS}

\begin{itemize}

\item FIGURE 1:\\
$\Sigma_{n,2}$ vs $k_{n}$ in log-log scale obtained from an integration of the 
model with $N=23$, $K_{0}=6.25\times10^{-2}$ 
and different values of the forcing:
$f=5\times10^{-3}(1+i)$ (plus), $f=6\times10^{-3}(1+i)$ (squares),
$f=8\times10^{-3}(1+i)$ (cross). 
Diamonds represent the results obtained in the
benchmark integration (i.e., the original GOY model). 

\item FIGURE 2:\\
$\Sigma_{n,6}$ vs $\Sigma_{n,3}$ in log-log scale obtained for the system
with the same parameters as in figure 1 and $f=8\times 10^{-3}(1+i)$ 
(diamonds). The K41 line (dashed line) is also shown for comparison.

\item FIGURE 3:\\
$\zeta_{p}$ vs p for the GOY model(diamonds) and for its reversible 
analog. Squares are obtained for $f=8\times10^{-3}(1+i)$
and pluses for $f= 5\times10^{-3}(1+i)$.
 The anomalous exponents have been calculated using the ESS.
\item FIGURE 4:\\
$C(\tau)$ vs $\tau$ for two different
values of the forcing: $f=5\times10^{-3}(1+i)$ (solid line) and
$f=8\times10^{-3}(1+i)$ (dashed line).

\end{itemize}

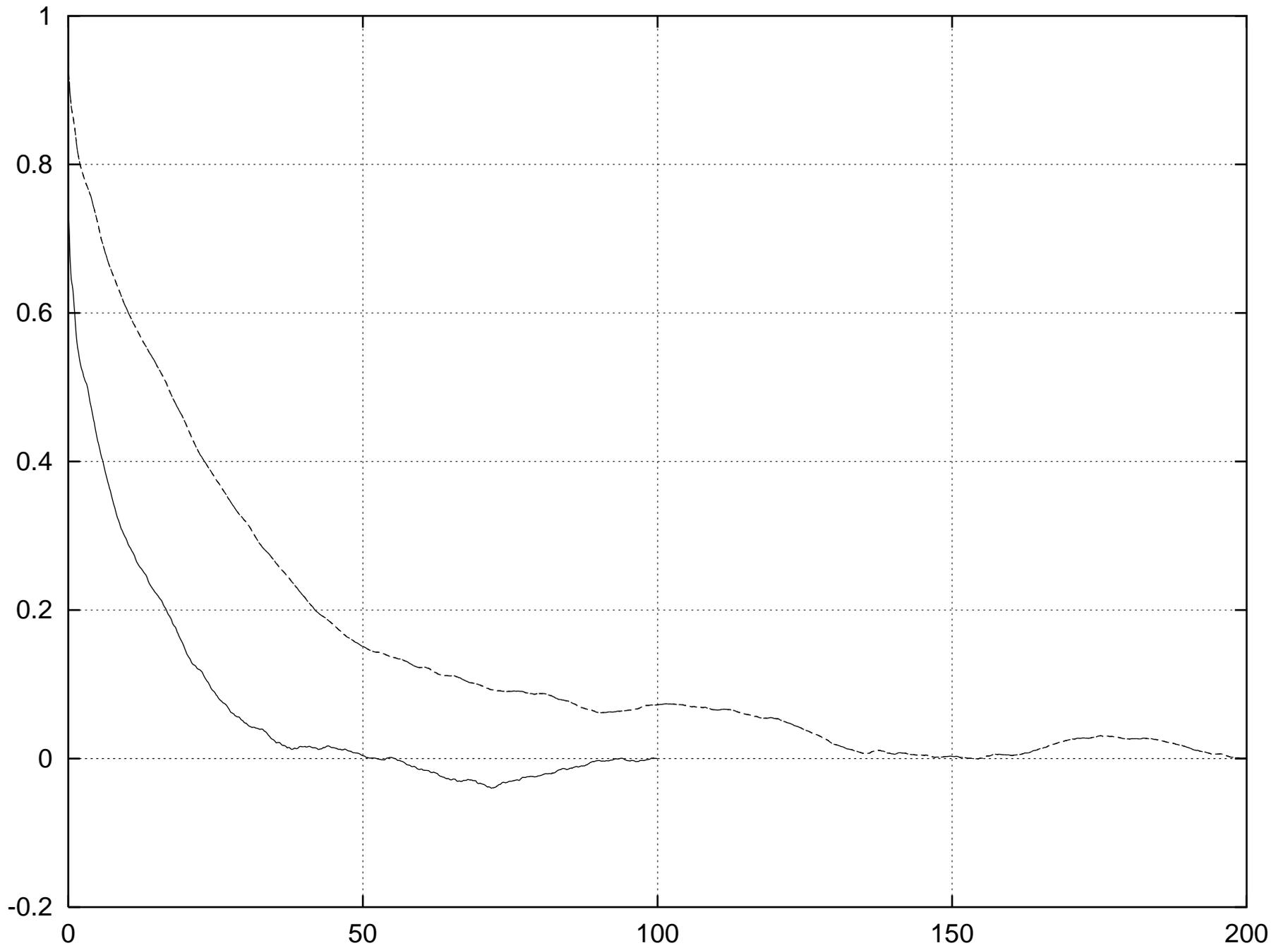
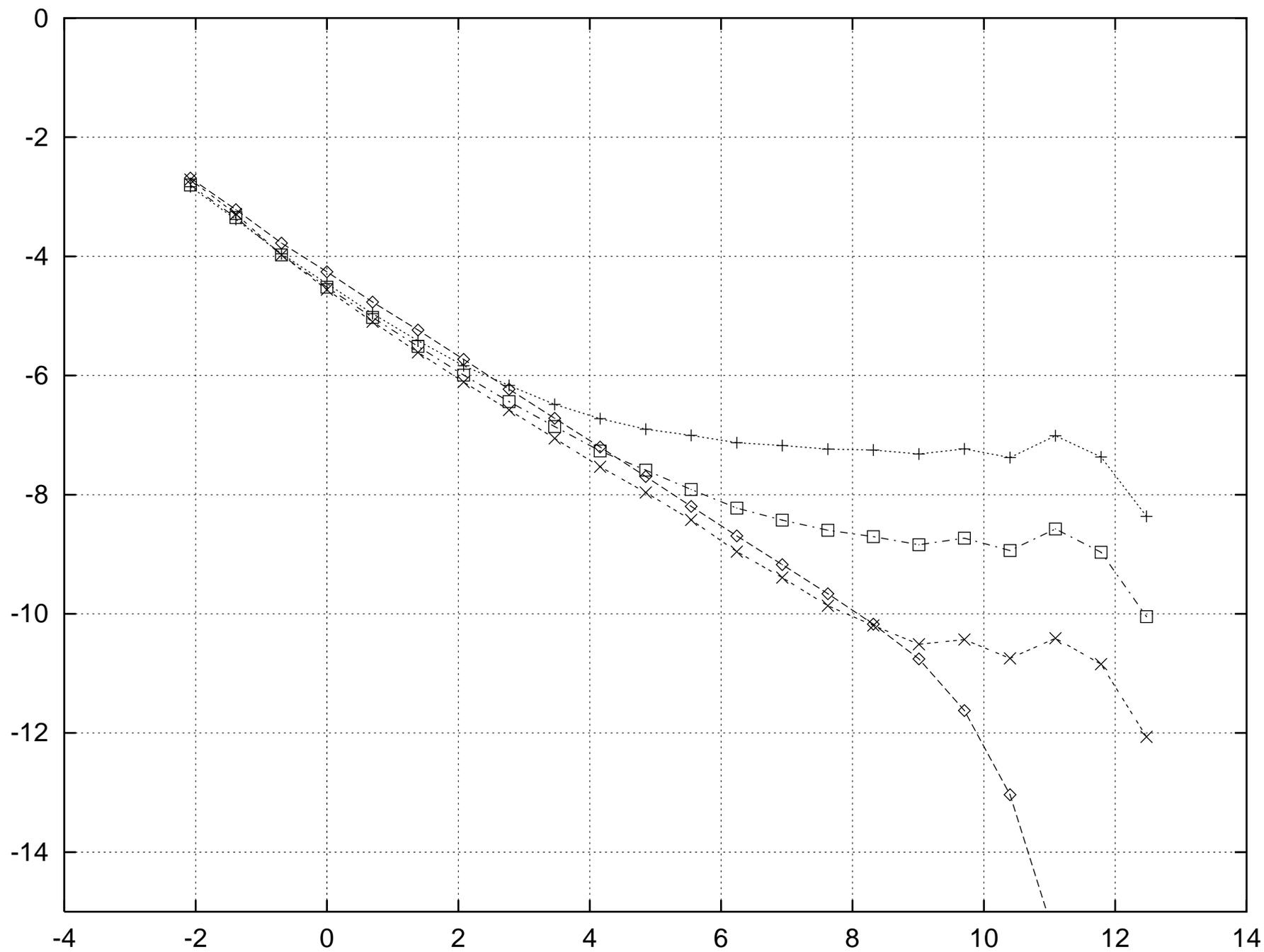
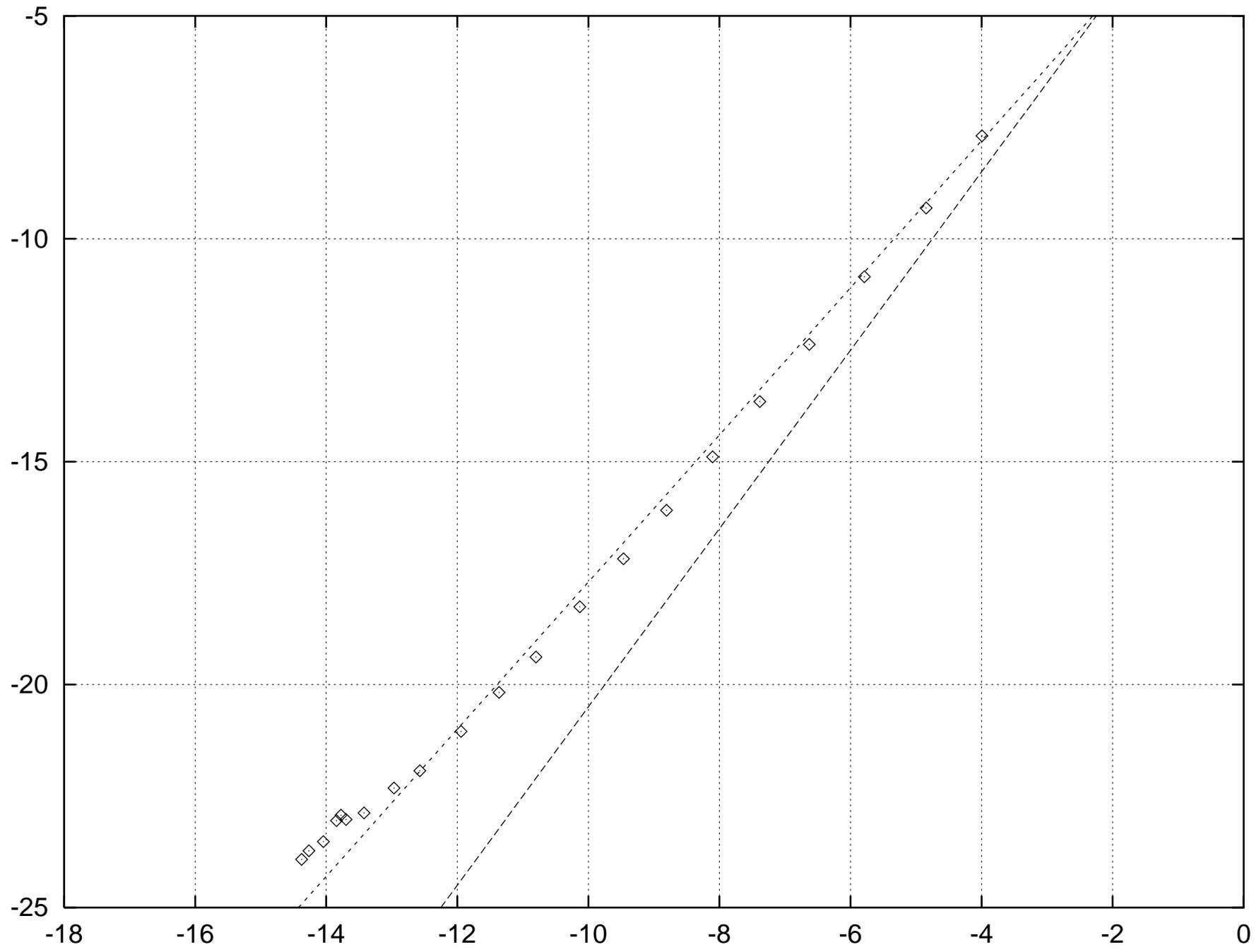
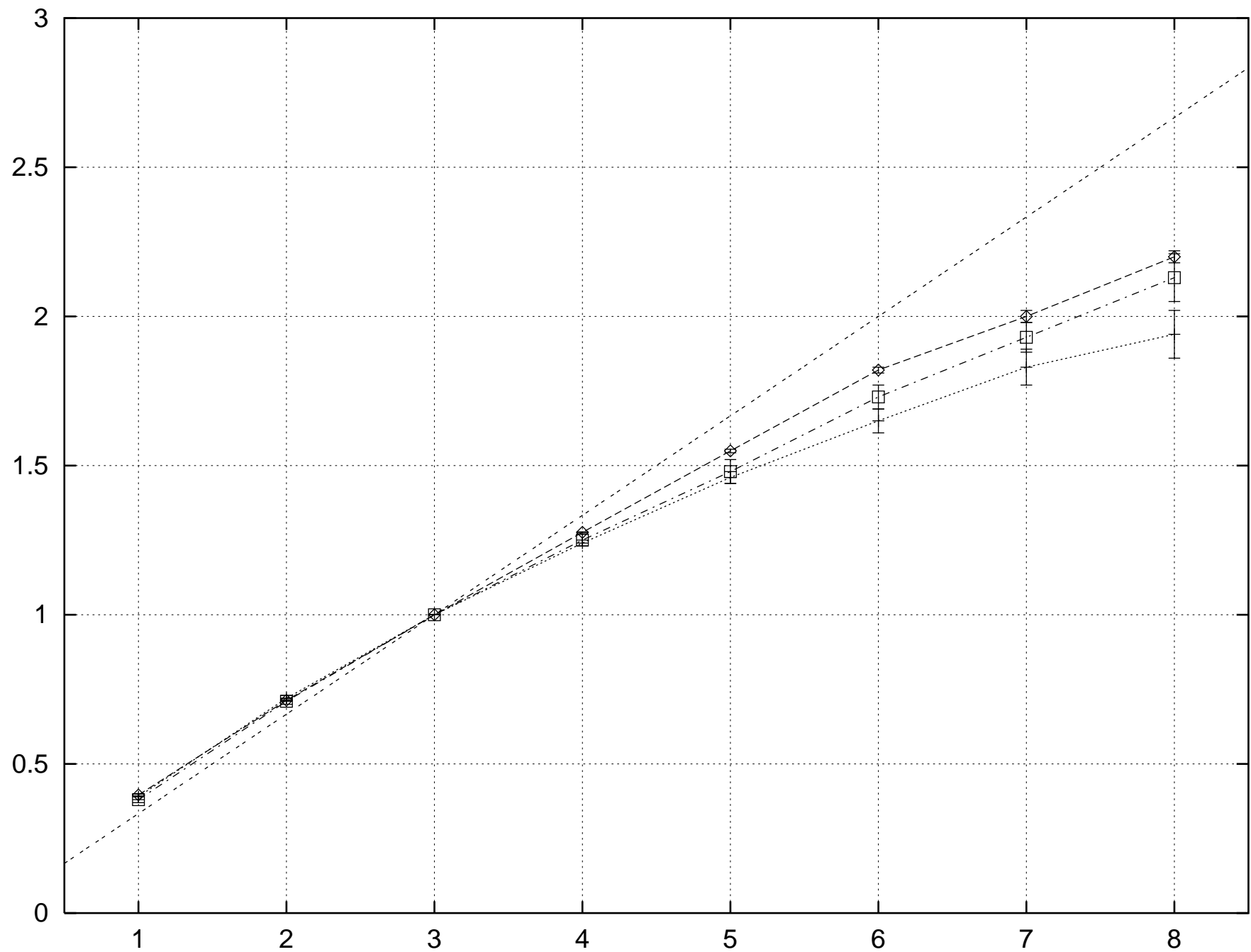
\end{document}